\documentclass[a4paper,12pt]{article}
\def\pTm   {\partial_{T_{aa'D'}}}

\def\ad   {a^{\dagger}}
\def\cd   {c^{\dagger}}

\def\bd   {b^{\dagger}}
\def\cd   {c^{\dagger}}

\def\bd   {\overline {D}}
\def\HP   {\hat {P}}

\def\al   {\alpha}

\def\HH {\hat H}

\def\HN {\hat N}

\def\HP {\hat P}
\def\HQ {\hat Q}
\def\HR {\hat R}
\def\Hrho  {\hat \rho}
\def\Hrhod  {\hat \rho^{\dagger}}
\def\HS {\hat S}
\def\HU {\hat U}

\def\HW {\hat W}

\def\HH {\hat H}

\def\HQ {\hat Q}

\def\HN {\hat N}
\def\HT {\hat T}

\def\CE {{\cal E}}

\def\CR {{\cal R}}

\def\CN {{\cal N}}
\def\CO {{\cal O}}

{\it}
{\it}
\def\expo {\rm {exp}}{\it}
\def\tr {\rm {tr}}{\it}
\def\Tr {\rm {Tr}}{\it}
\def\exp {\rm {e}}{\it}

\def\om {\omega}

\usepackage{times}
\usepackage{graphicx}
\begin{document}
\title{ Extension of the Time-Dependent Multi-Determinant approach to 
propagators.}
\author{G. Puddu\\
       Dipartimento di Fisica dell'Universita' di Milano,\\
       Via Celoria 16, I-20133 Milano, Italy}
\maketitle
\begin {abstract}
         We extend the recently proposed  Time-Dependent Multi-Determinant approach (ref.[1])
         to the description of fermionic propagators. The method hinges on equations of motion
         obtained using variational principles of Dirac type. In particular we study the trace
         of real and imaginary time propagators, i.e. the partition function. The method is equally 
         applicable
         with or without projectors to good quantum numbers. We discuss  as a numerical example
         the micro-canonical level density obtained from the propagation in real time. 
\par\noindent
{\bf{Pacs numbers}}: 21.60.-n,$\,\,$  24.10.Cn.
\par\noindent
{\bf{Keywords}}: Time-dependent variational principle; quantum dynamics,finite temperature, strongly correlated 
fermionic systems.
\vfill
\eject
\end{abstract}
\section{ Introduction.}
     In a recent work we introduced the Time-Dependent Multi-Determinant approach (ref.[1]) (TDMD),
     whereby the nuclear wave function is described as a linear combination of several Slater
     determinants. The Dirac variational principle has been used to obtain the equations of motion 
     describing the time evolution
     of the nuclear wave function from an initial state. The method has been  applied to the description
     of monopole resonances in a light nucleus. The TDMD has been shown to be a good dynamical
     approach in the sense that the initial energy and norm of the wave function are preserved by
     the equations of motion. In ref. [1] this has been shown without explicit reference to
     projectors to good quantum numbers (i.e. angular momentum and parity). The proof of 
     norm and energy conservation can be carried out also using projectors to good quantum
     numbers with only minor additional algebra on the assumption that the Hamiltonian 
     is rotational and parity invariant. The formalism of ref. [1]  also has been used for 
     imaginary time propagation which can be used to refine the ground state from an initial 
     wave function. The purpose of this work is to extend the formalism
     introduced in ref. [1] to the description of quantum mechanical propagators.
     The key ideas are the following. In ref. [1] the nuclear wave function is described as
$$
|\psi>=\sum_{S=1}^{N_w} |U_S>
\eqno(1)
$$
     where $|U_S>$ is a Slater determinant labeled by the index $S$ and each Slater
     determinant was written as 
$$
|U_S>= \cd_{1S}\cd_{2S}...\cd_{AS} |0>
\eqno(2)
$$
     $A$ being the number of particles, $S$ labels the Slater determinant and
$$
\cd_{\al S}= \sum_{i=1,N_s} U_{i,\al S } \ad_i,\;\;\;(\al=1,2,..,A)
\eqno(3)
$$
     are generalized creation operators written as a linear combination of the
     standard creation operators $\ad_i$ in the single-particle state $i$ (for
     example harmonic oscillator single-particle states). We can  conveniently
     recast each Slater determinant as
$$
|U_S>= \HU(S) \ad_1\ad_2..\ad_A |0>
\eqno(4)
$$
     where $\HU(S)$ is an operator of the type
$$
\HU(S)= \expo [ \ad u(S) a]
\eqno(5)
$$
     We use here a matrix notation, that is $\ad u(S) a= \sum_{ij}\ad_i u_{ij}(S) a_j$, where
     the sum runs over the single-particle states in the basis. We call the propagators
     of eq.(5) elementary propagators (EP).  The relation between
     the single-particle wave functions $U(S)$ and the matrix $u(S)$ is given by
$$
 U= \expo(u) 
\eqno(6)
$$
      Note that the matrix $u$ uniquely specifies the single-particle wave functions $U$,
      but the inverse statement is not true. In fact, in order to construct $U$ we need all matrix  
      elements of the matrix $u$, but only the matrix elements $U_{i,\al}$ for $i=1,2,..,N_s$ 
      and $\al=1,2,..,A$ are used in eq.(3). That is, only part of the 
      information contained in $u$ is used
      in the TDMD approach. We can state that
$$
|\psi>=\big[\sum_{S=1}^{N_w} \HU(S)\big] \ad_1\ad_2..\ad_A |0>
\eqno(7)
$$
      The sum in the square brackets is a propagator and the wave function is obtained 
      by projecting this propagator onto a reference state. Most of the information
      contained in the propagator is not used in the TDMD approach. It is natural to ask  
      what is the equation of motion for this propagator and what kind of information
      can be extracted from it.  We consider the following variational principles.
      If $\rho$ a propagator for a Hamiltonian $H$,  the propagator satisfies the
       following variational principles for real time,
$$
i Tr [\delta\rho^{\dagger} \dot \rho]= Tr [\delta\rho^{\dagger} H \rho]
\eqno(8a)
$$
$$
-i Tr [\dot\rho^{\dagger} \delta \rho]= Tr [\rho^{\dagger} H \delta\rho]
\eqno(8b)
$$
      and for imaginary time
$$
 Tr [\delta\rho^{\dagger} \dot \rho]= -Tr [\delta\rho^{\dagger} H \rho]
\eqno(9a)
$$
$$
 Tr [\dot\rho^{\dagger} \delta \rho]= -Tr [\rho^{\dagger} H \delta\rho]
\eqno(9b)
$$
     Eqs. (8a) and (9a) are called EOM1, and eqs. (8b) and (9b)	are called EOM2.
     It is trivial to show that these variational principles lead to
     the exact real time propagator and imaginary time propagator if
     we use the full Hilbert space. Once we solve the variational
     equations we can evaluate $\Tr \rho(t)$ and perform a Fourier analysis
     in the case of real time to obtain the micro-canonical level density, or extract the 
     free-energy in the case of imaginary
     time.
\par 
     These are the basic ideas of the extension of the TDMD method
     to propagators. That is, we use variational principles for the propagators
     under the assumption that they are written as a sum of time dependent
     elementary  propagators of the type $\expo( \ad u a )$.
\par
     We note that our method is very different from functional integrals
     based on the Hubbard-Stratonovich transformation (ref.[2]).
     With functional integrals
     we end up with multidimensional integrals of propagators,
     and they are hardly computable with Monte Carlo methods. The same can be said
     for imaginary time functional integrals, although in some cases the partition
     function is amenable to Monte Carlo evaluation.
     Moreover, in the case of the partition function our method is very different
     from the minimization of the free energy functional (see for example ref. [3]). In this
     latter method the free energy is a functional of a density operator
     (in some sense the propagator we have just described)  which
     must be minimized to obtain the actual propagator. The major stumbling block of
     this method is that we do not know how to compute the entropy in presence
     of a projector to good quantum numbers or if the density operator is a
     sum of elementary propagators. This difficulty has been the
     major hurdle in using  the temperature dependent Hartree-Fock, 
     or Hartree-Fock-Bogoliubov methods in presence of  exact projectors to good quantum numbers.
     This work deals mostly with formalism and only a few numerical examples.
\par
     The outline of this work is the following. In  section 2 we derive
     in detail the equations of motion. In section 3 we discuss a few properties
     in the imaginary time case. In section 4 we discuss the equations of motion for the
     real time propagation, conservation laws, and how one can extract the 
     micro-canonical level density from the  Fourier
     transform of real time propagators. In section 5 we discuss a few numerical examples
     in a simplified Hilbert space.
\bigskip
\bigskip
\section{ The equations of motion.}
\bigskip
\par
      As a notation, we use a caret to denote second quantized operators and
      we denote elementary propagators (EP) as
$$
\HU = \expo [\sum_{ij} \ad_i u_{ij} a_j]
\eqno(10)
$$
      where the sum runs over the single-particle space $i=1,2,..,N_s$. The matrix $u$ is
      time dependent. As well known, EP's  form a group (ref.[4]), and the product of
      two EP's is an EP. Also, to any EP we can associate the matrix (which we denote without the caret)
$$
U= \expo (u)
\eqno(11)
$$
      Throughout this work, small letters will denote the logarithm of matrices as in eq.(11)
      which are denoted with capital letters. In this work we do not consider the most general
      EP, where we allow for particle number violation as done in the Hartree-Fock-Bogoliubov
      formalism. The group property is still valid in such a case and it is the cornerstone
      of the method.
      Given two EP's such as $\HS$ and $\HT$ represented by $S$ and $T$ respectively, the product
      $\HT \HS$ is represented by $TS$. Moreover, traces taken in the full Hilbert space will be 
      denoted as $\Tr$ and traces taken in the single-particle space as $\tr$. 
      Since we shall consider traces in the Hilbert space,
      we start with Grand-Canonical traces, and we consider since the beginning, projectors
      to good quantum numbers, which we write schematically as
$$
\HP = \sum_E d^{\star}(E) \HR(E)
\eqno(12)
$$
      where $ \HR(E)$ is a rotation operator (which is an EP) dependent on the three
      Euler angles, in the case of the angular momentum projector, or is the operator
      $\expo (\alpha \HN)$, $\HN$ being the particle number operator, in the case of
      particle number projector. In this latter case $\alpha$ is a purely imaginary
      phase $ 2 i\pi k/N_s $ with $k=1,2,..,N_s$. Similarly we can recast the parity
      projector as in eq.(12). The detailed form of $d^{\star}(E)$ can be found in many textbooks
      (see for instance ref.[5]).
      The ansatz for the propagator is
$$
\hat\rho=\sum_{D=1}^{N_D} \HS_D
\eqno(13a)
$$
      and
$$
\Hrhod=\sum_{D'=1}^{N_D} \HT_{D'}
\eqno(13b)
$$
      we use this notation since often we  omit the labels $D$ and $D'$,
      in order to shorten the equations, with the understanding that $\HT_{D'}=\HS_{D'}^{\dagger}$.
      Consider the following traces
$$
\Tr [ \Hrhod \HP\Hrho] =\sum_{D'D} \sum_E d^{\star}(E) \Tr [\HS \HT \HR] 
\eqno(14)
$$     
      where, again, $\HT$ is labeled by $D'$ and $\HS$ by $D$. Note that we have used the cyclic 
      property of the trace.
      Using the aforementioned group property and the identity 
$$
\Tr \HW = \det(1+W)
\eqno(15)
$$
      valid for any EP, we have 
$$
\Tr [ \Hrhod\HP \Hrho] =\sum_{D'D} \sum_E d^{\star}(E)det [1+ STR]
\eqno(16)
$$
     To obtain the time derivative of $\Hrho$. Let us vary eq.(16) with respect
     to all $S$'s. Using the identity, valid for any matrix $M$,
$$
\delta \det M = \det M \tr(M^{-1}\delta M)
\eqno(17)
$$
      we obtain
$$
\Tr [ \Hrhod\HP \dot{\Hrho}] =\sum_{D'D} \sum_E d^{\star}(E)\det [1+ STR] \tr(F \dot S TR)
\eqno(18)
$$
     where 
$$
F = (1+STR)^{-1}
\eqno(19)
$$
      Next, we evaluate the variation of eq.(18) with respect to a specific $T_{aa'D'}$
      where $a,a'$ are single-particle indices.
      Again using the identity of eq.(17) we obtain
$$
\delta_{T_{D'}} \Tr [ \Hrhod\HP\dot{\Hrho}]=
\sum_D \sum_E d^{\star}(E) \det [1+ STR]\times
$$
$$
 [ \tr(R F S\delta T ) \tr(R F \dot S T)-
\tr(RF\dot S TRFS \delta T) +\tr(RF \dot S \delta T)]
\eqno(20)
$$
      Hence
$$
\pTm {\Tr [ \Hrhod\HP\dot{\Hrho}]}= \sum_D \sum_E d^{\star}(E) \det [1+ STR]\times
$$
$$
\big (  \tr(RF\dot ST)RFS - RF\dot S TRFS+ RF\dot S \big )_{a'a D'}
\eqno(21)
$$ 
      We now have to evaluate the right hand side of the variational equations.
      Let us assume that we have lumped together the kinetic energy and the two-body
      potential, as normally done in the shell model, and that
$$
\HH= {1\over 2}\sum_{ijkl} H_{ijkl}\ad_i\ad_j a_l a_k
\eqno(22)
$$
      where $H$ is already antisymmetrized (i.e. $H_{ijkl}=-H_{ijlk}$).
      We have 
$$
\Tr(\Hrhod \HP \HH \Hrho)= \sum_{D'D} \sum_E d^{\star}(E)\det [1+ STR]
[ \tr \Gamma \CN)]
\eqno(23)
$$
      where 
$$
\CN=1-F
\eqno(24)
$$
      and (the sum over repeated indices is assumed)
$$
\Gamma_{kl}=H_{ikjl} \CN_{ji}
\eqno(25)
$$
      We have then
$$
\delta_{T_{D'}} \Tr(\Hrhod \HP \HH \Hrho)= \sum_D \sum_E d^{\star}(E)\det [1+ STR]\times
$$
$$
[ \tr(\Gamma \CN)\tr(RFS \delta T)+2 \tr(RF\Gamma FS\delta T)]
\eqno(26)
$$
      Hence for a specified $T_{a a' D'}$ we have for the right hand side, which
      we call $\CR$,
$$
\CR^{(1)}_{a' a D'}= \sum_D \sum_E d^{\star}(E)\det [1+ STR]
[ \tr(\Gamma \CN) (RFS)+2(RF\Gamma FS)]_{a'a}
\eqno(27)
$$
     We are now in a condition to write down explicitly the equations of
     motion. Let us consider first  the imaginary time case.
     We have for $\dot S$ (the superscript refers to EOM1),
$$
L^{(1)}_{a'aD', b,b'D}\dot S_{bb'D}= - \CR^{(1)}_{a' a D'}
\eqno(28)
$$
     The matrix $L$ can be read off from eq. (21) and is given by
     the following expression
$$
L^{(1)}_{a'a D', b,b'D}= \sum_E d^{\star}(E)\det [1+ STR]\times
$$
$$
[(RFS)_{a'a} (TRF)_{b'b}- (RF)_{a'b}(TRFS)_{b'a}+\delta_{b'a}(RF)_{a'b}]
\eqno(29)
$$
      In the case of real time propagation the equation of motion EOM1 is
$$
L^{(1)}_{a'aD', b,b'D}\dot S_{bb'D}= -i \CR^{(1)}_{a'a D'}
\eqno(30)
$$
      The equations of motion EOM2 can be obtained in a similar way. We first 
      evaluate $ \Tr [ \dot{\Hrhod}\HP \Hrho]$ and then we vary the result with respect
      to $S_{a a' D}$. We simply write the result as
$$
L^{(2)}_{a'a D, b,b'D'}\dot T_{bb'D'}= i \CR^{(2)}_{a'aD}
\eqno(31)
$$
      for real time propagation and 
$$
L^{(2)}_{a'a D, b,b'D'}\dot T_{bb'D'}= - \CR^{(2)}_{a'aD}
\eqno(32)
$$
      in the case of  imaginary time. The matrices in eqs. (31) and (32) have the following
      expressions
$$
L^{(2)}_{a'aD, b,b'D'}= \sum_E d^{\star}(E)\det [1+ STR]\times
$$
$$
[(RFS)_{b'b} (TRF)_{a'a}- (RF)_{b'a}(TRFS)_{a'b}+\delta_{a'b}(RF)_{b'a}]
\eqno(33)
$$
     and
$$
\CR^{(2)}_{a' a D}= \sum_{D'} \sum_E d^{\star}(E)\det [1+ STR]
[ \tr(\Gamma \CN) (TRF)+2(TRF\Gamma F)]_{a'a}
\eqno(34)
$$
      In the case of real time EOM1 and EOM2 are the complex conjugate of each other.
\par
      If we consider neutrons and protons separately we simply have to add
      the extra isospin index to the single-particle indices. In such a case
      it is convenient to choose the matrices $S$ as block diagonal, i.e.
      the matrix $S$ does not couple neutrons and protons. This choice
      make  the projection to the proper number of neutrons and protons easier.
\bigskip
\bigskip
\section{ Some properties of the propagators for imaginary time.}
\bigskip
      Let us consider the imaginary time equations of motion. Let us assume
      that at some initial time $t_0$ the propagator is a sum of hermitian
      EP's and a sum of non-hermitian EP's plus their hermitian conjugates.
      We then can say that for each $\HS_D$ there is a $\bd$ (which can be
      $D$ itself) such that $S^{\dagger}_{a a'D} = S_{a a'\bd}$, or 
      $S^{\star}_{a'a D}=S_{a a'\bd}= T_{a a' D}$.
      Let us prove that the
      time evolution preserves the hermitian structure of $\Hrho$ and that
      EOM2 is equivalent to EOM1. These two properties are essential
      from a physical point of view. 
      In order to do so, let us set
$$
\CO=\Tr( \Hrhod \HP \Hrho),\;\;\;\; \CE=\Tr(\Hrhod \HP \HH\Hrho)
\eqno(35)
$$
      Since the projector satisfies the relations $\HP^{\dagger}=\HP$ and
      $\HP^2=\HP$ the above functional are real. Moreover we can rewrite
      EOM1 schematically as (using the sum convention)
$$
 {\partial^2 \CO\over\partial T_{aa'D}\partial S_{bb'D'}} \dot S_{bb'D'}=
 -{ \partial \CE\over \partial T_{aa'D}}
\eqno(36)
$$
      which can be rewritten as
$$
 {\partial^2 \CO\over\partial S_{aa'\bd}\partial S_{bb'D'}} \dot S_{bb'D'}=
 -{ \partial \CE\over \partial S_{aa'\bd}}
\eqno(37)
$$
      Taking the  complex conjugate of eq. (37) we have
$$
{\partial^2 \CO\over\partial S_{a'a D}\partial S_{b'b\bd'} } (\dot S_{bb'D'})^{\star}= -{ \partial \CE\over
 \partial S_{a'aD}}
\eqno(38)
$$
       Hence, comparing eqs. (37) and (38) we have
$$
(\dot S_{bb'D'})^{\star}= \dot S_{b'b\bd'}
\eqno(39)
$$
       that is $ \dot S_{D'}= \dot S^{\dagger}_{\bd'}$.
       This implies that 
$$
 (S+dt \dot S)^{\dagger}_{D} = (S+dt \dot S)_{\bd}
\eqno(40)
$$
       Strictly this is true if the solution of the system of eq.(37) is unique.
       If we have multipole solutions we can always force eq.(38) and still satisfy
       eq.(37). This property guarantees that the spectrum of $\Hrho$ is real.
\par
       Next, EOM2 can be rewritten as
$$
 {\partial^2 \CO\over\partial S_{aa'D}\partial T_{bb'D'}} \dot T_{bb'D'}=
 -{ \partial \CE\over \partial S_{aa'D}}
\eqno(41)
$$
       which can be recast as
$$
 {\partial^2 \CO\over\partial S_{aa'D}\partial S_{bb'\bd'}} \dot T_{bb'D'}=
 -{ \partial \CE\over \partial S_{aa'D}}
\eqno(42)
$$
       which implies that $\dot T_{bb'D'}= \dot S_{bb'\bd'}$. Therefore
       EOM2 gives the same solution as EOM1.
\bigskip
\bigskip
\section{ The propagators for real time.}
\bigskip
\par
{\it{3a. Micro-canonical level density and constants of motion.}}
\par
     The equations discussed in the previous section are not easy to solve.
     In fact, we expect that the numerical solution will show an exponential
     behavior as a function of the imaginary time and therefore some
     kind of numerical stabilization might be necessary especially for
     large values of imaginary time. In this section we discuss the real time
     propagation and a  motivation on physical grounds.
\par
     Let us assume that we have solved EOM1 (or EOM2) as a function of the time
     and let us evaluate 
$$
f(t)= \Tr \Hrho(t)
\eqno(43)
$$
     Consider the following Fourier transform
$$
g(\om) = {1 \over \pi} Re \int_0^{-\infty} dt f(t)\exp^{i(\om+i\gamma)t}
\eqno(44)
$$
     in the limit of $\gamma\rightarrow 0^+$. The function $g(\om)$ 
     approaches the level density if we have evaluated $f(t)$ with
     sufficient accuracy for the Hamiltonian $\HH$. There are a number of points to 
     be discussed. Consider first some conservation laws which must be satisfied
     if we have solved the equations of motion accurately.
     We shall prove that the quantities defined in eq.(35) are constants in time.
     Let us rewrite EOM1 and EOM2 in the following form
$$
 {\partial^2 \CO\over\partial S^{\star}_{aa'D}\partial S_{bb'D'}} \dot S_{bb'D'}=
 -i{ \partial \CE\over \partial S^{\star}_{aa'D}}
\eqno(45)
$$
     and
$$
 {\partial^2 \CO\over\partial S_{aa'D}\partial S^{\star}_{bb'D'}} \dot S^{\star}_{bb'D'}=
 i{ \partial \CE\over \partial S_{aa'D}}
\eqno(46)
$$
     Multiplying eq.(45) by $ \dot S^{\star}_{b b' D}$ and summing over the indices,
     multiplying eq.(46) by $\dot S_{a	a' D'}$ and summing over the indices, and
     subtracting the two results one has
$$
{\partial \CE\over \partial S^{\star}_{aa'D}}\dot S^{\star}_{b	b' D}+
{\partial \CE\over \partial S_{aa'D}} \dot S_{a  a' D'}=0
\eqno(47)
$$
     The above is the time derivative of $\CE$.
\par
     The conservation of $\CO$ is slightly more involved to prove. Consider EOM1 as given by
     eq.(30) and EOM2 given by eq.(31). Let us multiply EOM1 by $T_{a a' D'}$ and sum over the 
     indices, and EOM2 by $S_{aa'D}$ and sum over the indices and add the two results. 
     We obtain, using the cyclic property of the trace and the definitions of $F$ and $\CN$
$$
{d \over dt}\sum_E  d^{\star}(E)  \sum_{DD'} [ \det(1+STR) \tr (\CN) ] =0
\eqno(48)
$$
     Let us now consider separately the particle number projection from the rest
     of the projectors to good quantum numbers, in the following way.
     Let us define the complex fugacity $z=\expo(\alpha)$ and isolate it from the rest of
     the rotation operator. Then
$$
{d \over dt}\sum_E  d^{\star}(E)  \sum_{DD'} [ \det(1+z STR) \tr (\CN) ] =0
\eqno(49)
$$ 
    and 
$$
\CN= zSTR/(1+zSTR)
\eqno(50)
$$
     The exact projection to the proper number of particles $A$ can be done by isolating
     the coefficient of $z^A$ in $ \det(1+z STR) \tr (\CN)$. Let us consider the diagonal 
     representation of the matrix $W=STR$ and let us call $\om_{\mu}$ its eigenvalues.
     Recall (although we work in real time) that $\det(1+z STR)$ is a grand-canonical
     partition function and that the canonical partition function for $A$ particles is
     given by
$$
C(A)= \sum_{\mu_1<\mu_2<...\mu_A} \om_{\mu_1}\om_{\mu_2}...\om_{\mu_A}
\eqno(51)
$$
     This is a homogeneous polynomial of power $A$ in the $\om$'s for which the Euler's
     theorem holds. Let us call $C(A-1,\mu)$ the canonical partition function for
     $A-1$ particles with the level $\om_{\mu}$ removed. Then eq. (49) can be rewritten as
$$
{d \over dt}\sum_E  d^{\star}(E)  \sum_{DD'} \sum_{\mu}\om_{\mu}C(A-1,\mu)= 0
\eqno(52)
$$
     Since $C(A-1,\mu)=\partial C(A)/\partial \om_{\mu}$ the Euler's theorem gives
     $\sum_{\mu}\om_{\mu}C(A-1,\mu)=A C(A)$, and therefore
$$
{d \over dt}\sum_E  d^{\star}(E)  \sum_{DD'} C(A) = 0
\eqno(53)
$$
     hence the particle-number projected overlap is a constant of motion.
     These two conservation laws are a valuable test in order to control the accuracy
     of the time evolution. 
     There are a few remaining points which will discussed in the next subsection.
\par
{\it{3b. The choice of the intial conditions.}}
\par
      We have described in detail
     the form of the equations of motion but so far we have not specified the initial condition
     at $t=0$. Ideally we would set $\rho=1$. This choice is necessary if we wish to 
     evaluate the micro-canonical level density using eq.(44). Note however
     that we are solving an initial value problem and in principle we can take any
     initial $\rho(0)$. If we consider $\rho(0)=1$, we can only consider one single 
     EP. There is simply no way to have $\rho(0)=1$ with several independent EP's.
     In the case of several EP's we have several choices. Consider for a moment
     the decomposition of the Hamiltonian $\HH$ into a sum of quadratic
     operators of the type
$$
\HH= \hat h -\sum \HQ^2
\eqno(54)
$$
     much is the same way it is done as a preliminary step to express the propagator
     with functional integrals. In eq.(54) $\hat h$ and $\HQ$ are one body operators.
     The propagator after a small time interval $\delta t$, up to $\delta t^2$ terms,
     can be be written as 
$$
\rho(\delta t) = \expo (-i \delta t \hat h) +1/2 \sum [ \expo( i \sqrt{\delta t} \HQ)+
  \expo (- i \sqrt{\delta t} \HQ)]
\eqno(55)
$$
     As an initial start we can consider few terms of this type. In practice we do the following instead.
     Consider simply a sum of the type
$$
\rho(\delta t) = \sum \expo (-i \delta t \hat s)
\eqno(56)
$$
     where $\hat s$ are one body operators, unspecified for the moment. For sufficiently
     small $\delta t$, only their sum contributes to the propagator, that is,
     the ansatz of eq.(56) is equivalent to the choice of just one EP. Hence we first
     start from $\rho(0)=1$, using only one EP. We solve up to $\delta t$ the equations of motion
     and we decompose $S(\delta t)$ into a sum of different EP's. Such a sum of independent
     $S_D$ is our choice for the initial start. The set of $S_D$ is then evolved up to finite times.
     Since the decomposition of the initial $S$ (for $N_D=1$) into several $S$'s is arbitrary, we expect the
     the solution of the equations of motion is not unique. That is the matrix $ L$ can have
     $0$ eigenvalues. We test the eigenvalues of $L$ and we solve the linear system of equation
     (45) in the unknowns $\dot S_{bb'D'}$ in the basis that diagonalizes $L$. In doing so, we
     discard all $0$ eigenvalues and reconstruct $\dot S_{bb'D'}$ in the original basis.
     That is we use the generalized inverse of $L$.
     As a consistency test we verify that the $\dot S_{bb'D'}$ obtained in this way satisfy
     the original system of eq.(45). 
     An additional choice is to write $\rho(0)$ as a sum of EP's such that for small $\delta t$
     $\rho(0)$ is proportional to $1$ up to $\delta t^2$ terms.
     Note that in general especially for large
     single-particle spaces, the sizes of the linear system to be solved for $\dot S_{bb'D'}$
     can be very large and mathematical libraries such as SCALAPACK (ref.[6]) that can distribute
     large matrices into several processors are necessary. The time evolution of $\rho$ is obtained
     using Runge-Kutta methods of high accuracy. We give a few examples of the numerical solution
     of the equations of motion in the next section. As a final remark, we found that 
     even if our initial start for $\rho$ is unitary, unitarity is broken as we evolve at finite times.
     This raises the question  whether the number of levels obtained from eq.(44) is the correct one.
     We do not have in a strong argument regarding this point. However we can state that
     the integral over the energy  of the micro-canonical level density has the correct value.
     The argument is the following. Consider first the case of one EP. 
     The projected overlap at $t=0$ is simply the projected trace of the unity operator. Hence
     it is simply the total number of levels. Such an overlap is a constant of motion, even if
     unitarity is broken at finite times. In the case of several EP's, since at the initial time
     we decompose the propagator obtained after a small time interval $\delta t$ into several EP's,
     the energy integrated micro-canonical level density is the same (up to $\delta t^2$ terms) 
     and, again, after we solve  the equations of motion at finite time, we obtain
     approximately the correct value. 
     
\bigskip
\bigskip
\section{ A numerical example.}
\bigskip
\bigskip
     Let us consider a system of 6 neutrons in the 1s1p harmonic oscillator shells. We choose 
     the harmonic oscillator frequency $\hbar\Omega=12 MeV$. For the interaction we take
     the neutron-neutron part of  the N3LO interaction (ref.[7]) renormalized to the above
     single-particle space. This model is highly schematic and it serves solely to the
     purpose of testing the numerical method and the concepts of the previous sections.
     Since we solve the equations of motion in real time we must ensure to have the proper number of
     particles. We cannot use chemical potentials as usually done in the case of the imaginary
     time propagation. We also implement an angular momentum projector to $J_z=0$. In one
     case we use the full angular momentum projector to $J=0$. 
     This model has only $10$ states with $J_z=0$ and $4$ states with $J=0$. The full space 
     contains $28$ states. In all calculations we take $\delta t$ in the range of $10^{-4}\div 10^{-5}$.
     Let us consider first one elementary propagator, that is $N_D=1$.
     We start from $\rho(0)=1$.
\par
\renewcommand{\baselinestretch}{1}
\begin{figure}
\centering
\includegraphics[width=10.0cm,height=10.0cm,angle=0]{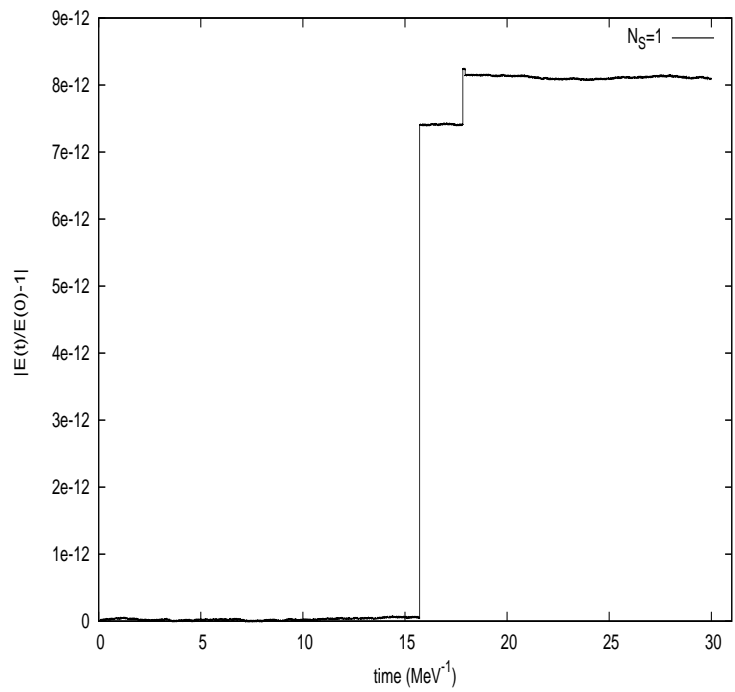}
\caption{Variation  $|\CE(t)/\CE(0)-1|$ as a function of time.}
\end{figure}
\renewcommand{\baselinestretch}{2}
     In fig.1 we show the error in the  conservation of the energy-like quantity $\CE(t)$ defined
     in eq.(35). Throughout this work we use $\hbar=1$, that is, we measure the time in units of $MeV^{-1}$.
     In fig.2 we show the deviation from unitarity. As it can be seen, although the propagator is not
     unitary, $ \CE(t)$ is constant for very long times. In fig. 3 we show the micro-canonical 
     level density given by eq. (44) as a function of the energy, together with the
     number function
$$
n(E) =  \int_{\infty}^E dE' f(E')
\eqno(57)
$$
\par
\renewcommand{\baselinestretch}{1}
\begin{figure}
\centering
\includegraphics[width=10.0cm,height=10.0cm,angle=0]{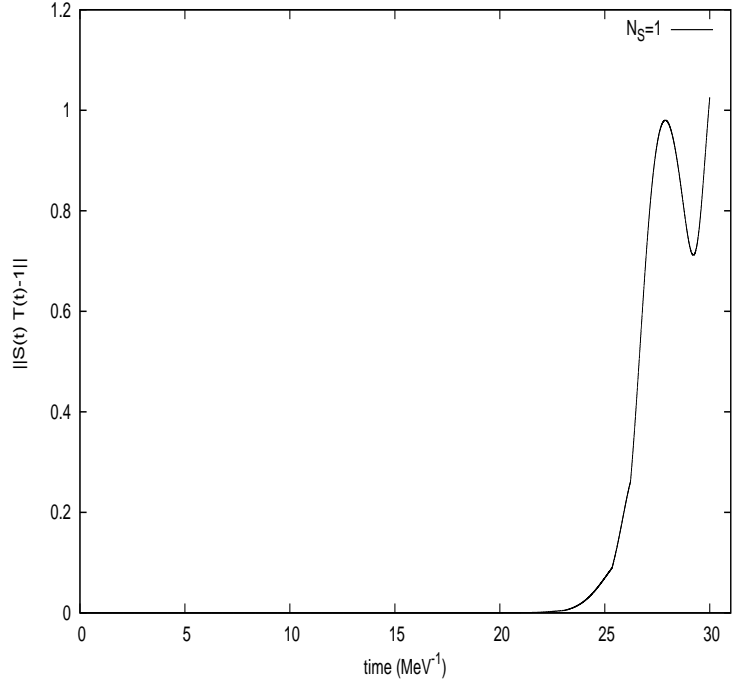}
\caption{Deviation from unitarity $||T(t) S(t)-1||$ as a function of time.}
\end{figure}
\renewcommand{\baselinestretch}{2}
     The number function counts the levels from $-\infty$ up to a given energy $E$.
     We took in eq.(44) $\gamma=0.1MeV$. In the limit $\gamma=\rightarrow 0$, $f(E)$ is
     a sum of Dirac-delta functions and the number function increases by one unit
     anytime we cross a level. As it can be seen from fig.3, in some cases $n(E)$ increases
     by two units, which points out to a degeneracy (or near degeneracy) of two levels,
     not separated by $\gamma=0.1MeV$. 
     as a check, note that the total number of levels is the correct one.
\par
\renewcommand{\baselinestretch}{1}
\begin{figure}
\centering
\includegraphics[width=10.0cm,height=10.0cm,angle=0]{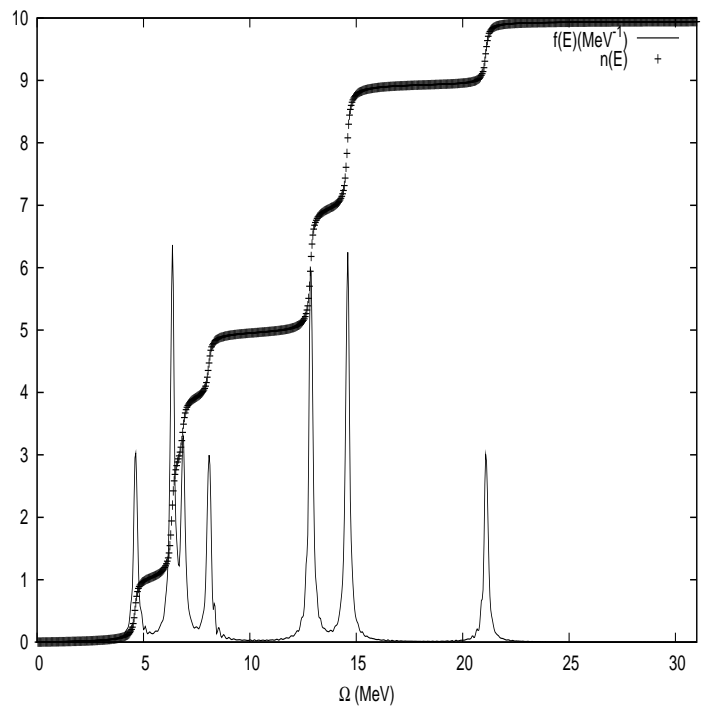}
\caption{ Level density $f(E)$ and number function $n(E)$ for $J_z=0$.}
\end{figure}
\renewcommand{\baselinestretch}{2}
     In fig.4 we show the level density and the number function for $N_D=1$ using
     the full projector to $J=0$. Note that the projector gives a different
     approximation to the full propagator compared to the $J_z=0$ case.
\par
\renewcommand{\baselinestretch}{1}
\begin{figure}
\centering
\includegraphics[width=10.0cm,height=10.0cm,angle=0]{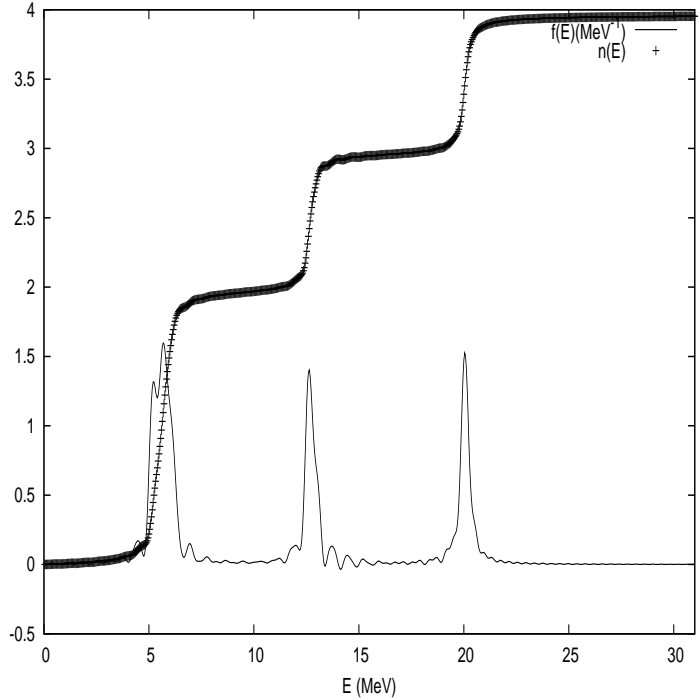}
\caption{ Level density $f(E)$ and number function $n(E)$ for $J=0$}
\end{figure}
\renewcommand{\baselinestretch}{2}
\par
          Although very schematic and simple, these two examples
         show the main features of the formalism and properties 
         described in the previous sections.
\par
     In conclusion, we have generalized the time dependent 
     multi-determinant approach
     to propagators using variational principles of Dirac-type.
     We described in detail the equations of motion and showed
     that there are constants of motion not related to the
     unitarity of the propagator. Such constants of motion are
     very useful to test the correctness and accuracy of the numerical methods.
     In the future we plan to extend these numerical techniques to the
     neutron-proton case for reasonably large shell model spaces.

\vfill
\eject
\end{document}